\documentclass[article,aps,pra,twocolumn]{revtex4}  

\usepackage{amsmath}    
\usepackage{graphicx}   
\newcommand{\ket}[1]{\left|#1\right\rangle}
\newcommand{\0}{\ket{0}}
\newcommand{\1}{\ket{1}}

\begin{document}

\title{Quantum spread spectrum multiple access}
\author{Juan Carlos Garcia-Escartin}
 \email{juagar@tel.uva.es}  
\affiliation{Universidad de Valladolid, Dpto. Teor\'ia de la Se\~{n}al e Ing. Telem\'atica, Paseo Bel\'en n$^o$ 15, 47011 Valladolid, Spain}
\author{Pedro Chamorro-Posada}
\affiliation{Universidad de Valladolid, Dpto. Teor\'ia de la Se\~{n}al e Ing. Telem\'atica, Paseo Bel\'en n$^o$ 15, 47011 Valladolid, Spain}
\date{\today}

\begin{abstract}
We describe a quantum multiple access scheme that can take separate single photon channels and combine them in the same path. We propose an add-drop multiplexer that can insert or extract a single photon into an optical fibre carrying the qubits of all the other users. The system follows the principle of code division multiple access, a spread spectrum technique widely used in cellular networks.
\end{abstract}
\maketitle

\section{Multiple access in photonic communication channels}
Quantum communication protocols such as quantum key distribution \cite{BB84,Eke91}, dense coding \cite{BW92} or quantum teleportation \cite{BBC93} expand the possibilities of classical data transmission. The quantum states of light are the natural choice to implement most of these protocols and there are successful optical implementations of quantum key distribution systems that send single photons through commercial optical fibre links \cite{GRT02}.

As the number of users grows, there appears the problem of channel access. Many different users might want to use the same resources at the same time. Classical communication networks solve this problem with a variety of multiple access schemes \cite{Skl83,Skl01}. Quantum networks have already made use of wavelength and frequency division multiple access techniques where each user sends data at different frequencies \cite{BBG03,OC06,YFT12,MRA12,CMP14,PDL14}, as well as time division multiple access where users wait for their turn \cite{CYT11}. In our previous work, we have also proposed different multiple access schemes using coherent states \cite{GC09} and the orbital angular momentum of single photons \cite{GC08,GC12}. 

In this paper, we propose a multiple access scheme based on spread spectrum techniques. The scheme is described for optical fibre, but could be easily adapted to free-space transmission. Our system is built from widely used optical elements. It sends the photons of multiple users through the same optical fibre so that they share their path, frequency band and time window. Previous spread spectrum multiple access techniques for quantum optical communication introduce heavy losses when they combine and separate the data from each user \cite{HUM11,Raz12,ZLO13}. Our scheme can, in principle, achieve almost deterministic operation. It follows an add-drop architecture with simple combination and extraction points and it has been designed to adapt classical spread spectrum methods directly. The system can reuse existing code families and most of the classical techniques. The users only need to add the multiplexer and demultiplexer systems we describe.

\section{Spread Spectrum} 
In cellular communication networks, spread spectrum techniques are often used to share the channel. In spread spectrum, a modulated data signal $d(t)$ of bandwidth $W$ is transformed so that it becomes a spread signal $s(t)$ of a larger bandwidth $SW$, where $S$ is called the spreading factor \cite{PSM82}. 

We are going to discuss a method based on direct-sequence spread spectrum technologies and their application to Code Division Multiple Access, CDMA. In CDMA, each user $U_i$ from a group of $N$ users, $U_1, U_2 \ldots, U_N$, is assigned a code $c_i$. We describe codes $c_i$ as vectors of elements $1$ and $-1$. The codes are chosen to be orthogonal, with $c_i\cdot c_j^T=\delta_{ij}$, or nearly orthogonal, with $c_i\cdot c_j^T\leq m$ for $i\neq j$ and an integer value of $m$ as small as possible.

\begin{figure}[htbp]
\centerline{\includegraphics{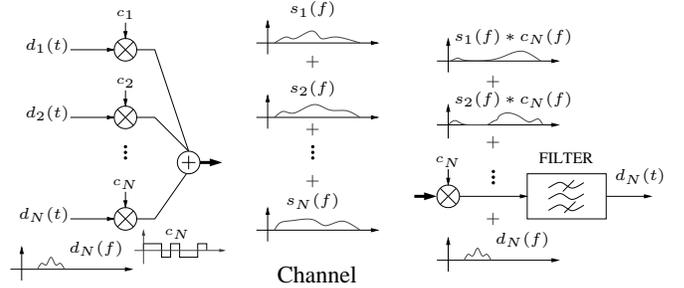}}
\caption{Example of a direct-sequence spread spectrum multiple access system. $N$ data signals $d_i(t)$ are multiplied by their corresponding $c_i$ code and sent together through the same channel. At the receiver, we can recover any desired data signal. For instance, if we want $d_N(t)$, we can multiply the received total signal by code $c_N$. Data signal $d_N(t)$ will return to its original region of the spectrum, but all the other signals will remain spread. If the codes are fully orthogonal, a band-pass filter will cancel the contribution of all the other channels, but even if they aren't, the energy of the interfering signals is spread through a bandwidth $SW$ and only a fraction $1/S$ will affect the desired signal.\label{spread}} 
\end{figure}

Figure \ref{spread} shows the spreading and despreading processes. We can basically consider spreading as a second modulation where the signal is multiplied by $c_i$. Spreading can be undone at the receiver if we multiply again the received signal and $c_i$.

If the codes are chosen well, the signals from all the users can be clearly separated even though they have shared the whole bandwidth at the same time. Furthermore, spreading provides improved results against noise and allows to increase the number of users beyond the perfect separation limit given by the finite number of orthogonal codes of a certain length. In spread spectrum there is a ``gentle degradation'' where each additional user above the limit appears only as a low level noise. 

Direct-sequence spread spectrum methods can be directly applied to single photons \cite{BCY10}. The photon's wavefunction can be spread through an extended bandwidth and later be recovered at the receiver. The despreading procedure takes the wavefunction back to its original bandwidth and, at the same time, spreads the noise which is then filtered. 

In this paper, we show how to take advantage of single photon spreading in a multiple access system for photonic channels.

\section{Building blocks}
Our system uses three standard optical fibre elements: optical modulators, circulators and fibre Bragg gratings. Electro-optic modulators alter the waveform of a signal according to a control signal \cite{ST91}. This operation can be extended to the quantum regime \cite{CF10}. We require an optical modulator that acts on the photon's phase. We can follow \cite{BCY10} and introduce a phase shift $\pm\frac{\pi}{2}$ to different time portions of the photon's wavefunction. A wavefunction of time length $T$ can be divided into $S$ segments. Each element of the code $c_i$ decides whether the modulator applies a phase shift $\frac{\pi}{2}$ (if the element corresponding to our time segment is 1) or $-\frac{\pi}{2}$ (if it is -1). We can also use an optical modulator which introduces a phase 0 (for an element 1) or a phase $\pi$ (for -1). The total effect is the same for both modulations, $\pm\frac{\pi}{2}$ or $0/\pi$, except for an unimportant global phase $\frac{\pi}{2}$.

\begin{figure}[htbp]
\centerline{\includegraphics{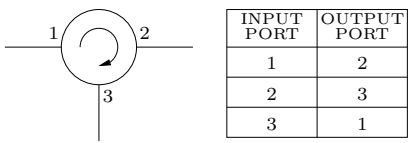}}
\caption{Symbol of the circulator and the correspondence between input and output ports. The element is not reciprocal. Light entering port 1 goes out port 2, but light coming into port 2 moves on to port 3 instead of going back to port 1.\label{circulator}} 
\end{figure}

We now need a method to combine the spread photons of all the users in the same optical fibre. We use two elements: circulators and fibre Bragg gratings, FBG. Circulators are non-reciprocal optical three- (or more) port devices that reroute incoming signals to the next output port (see Figure \ref{circulator}).

\begin{figure}[htbp]
\centerline{\includegraphics{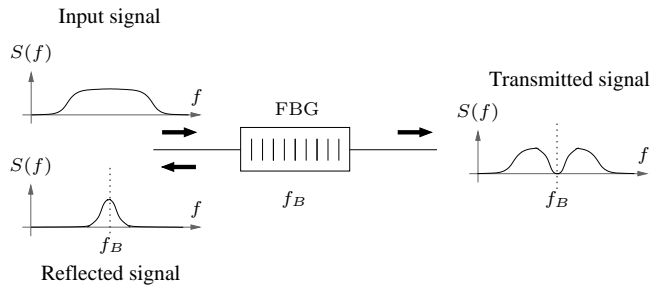}}
\caption{Fibre Bragg grating. The power spectral density of an incoming signal is divided into two parts. The grating acts as a band-stop filter that reflects the part of the spectrum around a frequency $f_B$ and transmits the rest of the input signal.\label{FBG}} 
\end{figure}

Fibre Bragg gratings are frequency specific reflectors that, in their most common form, let most of the incoming signal pass unaffected while a specific frequency band is reflected (see Figure \ref{FBG}).

\section{Signals}
Our data signals are qubits that can have different encodings. Frequency modulation is common in classical spread spectrum systems. We can follow that model and use frequency qubits, where states $\ket{0}$ and $\ket{1}$ correspond to wavefunctions at different frequencies \cite{CF12} or use phase modulation on the sidebands of a strong carrier \cite{GMS03}. 

Alternatively, we can use time-bin encoding and send a photon with a wavefunction restricted to a time window $(0,T_0)$ to encode state $\ket{0}$ or introduce a delay to move the wavefunction to time window $(T_0,2T_0)$ to encode state $\ket{1}$ \cite{BGT99}. Time-bin qubits are particularly attractive for optical fibre transmission and this is the preferred encoding in many experimental quantum key distribution systems. In this case, the codes must be designed for a time period of length $T_0$ and be applied twice. Otherwise, two orthogonal codes with the same first or second half could produce the same spread signals for different users and the signals can interfere.

\section{Multiplexing}
\begin{figure}[htbp]
\centering
\includegraphics{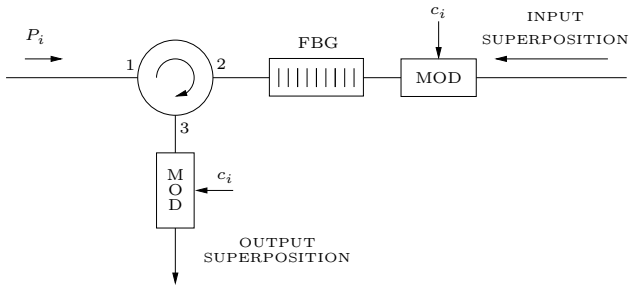}
\caption{Multiplexer: The new user $U_i$ wants to add photon $P_i$ to the input superposition coming from the common optical fibre. The central element is the fibre Bragg grating where the user's photon $P_i$ and the input superposition meet. Photon $P_i$ comes from port 1 of the circulator. The input superposition has been previously multiplied by $c_i$. The grating reflects the band of the spectrum which contains the whole wavefunction of $P_i$. The photons of the input superposition only have a fraction $1/S$ of their wavefunctions in that part of the spectrum. The joint signal which contains the new and old photons comes into port 2 of the circulator and goes out port 3 where a second modulator restores the old photons to their original state and spreads $P_i$. The new superposition is then put into the same optical fibre.\label{MUX}} 
\end{figure}

First, we discuss how the transmitter of each user can add its signal to a superposition state $\ket{\psi_S}$ that carries the photons of all the previous users. Figure \ref{MUX} shows the block diagram of the multiplexer. The multiplexing and demultiplexing systems follow the ideas of add-drop multiplexers that are extensively used in optical fibre networks to combine channels of different frequencies. We use them to combine single photons.

Most classical methods that add two signals do no work well with individual photon states. Quantum operations must be reversible. Two photons cannot be directly combined using common classical devices such as Y-junctions, which take two inputs into the same output. Irreversibility introduces a form of measurement that destroys quantum coherence. There are reversible couplers, but existing classical methods for optical CDMA \cite{Sal89,FM07} use elements like star couplers that give a part of the signal to all of the users and introduce high losses that do not scale well for single photons. The same problem limits previous quantum spread spectrum multiple access methods \cite{HUM11,Raz12,ZLO13}. We propose an architecture that minimizes photon loss. 

In our system, new photons are added in the multiplexer of Figure \ref{MUX}. When the input superposition reaches user $U_i$, it is multiplied by code $c_i$ with a modulator. The resulting signal is then directed to a fibre Bragg grating which reflects the frequency band that contains the photon signal before spreading.

At the same time, we send the modulated, but not yet spread, data signal $d_i(t)$ of the new user (photon $P_i$) through a circulator so that it reaches the FBG at the same time as the superposition signal but in the opposite direction. The signal that comes out of the FBG from the left and goes to the circulator includes the new photon and most of the spread superposition. The grating reflects the new photon into port 2 and a residual part of the spread superposition back to the fibre it came from. 

The signal with the old photons and the new photon reaches the circulator and is directed to a second modulator in port 3 which applies once more the code $c_i$. The input superposition is now back to its original form (except for a small loss) and the photon from user $U_i$ is spread. The resulting output is a superposition of the previous wavefunctions plus the new photon. The photons do not interact. The new term can be though of as the tensor product of orthogonal wavefunctions sharing the same frequency band. 

The fibre Bragg grating reflects a small part of the wavefunction of the incoming photons. If the spreading factor $S$ is large enough, the probability that a photon is reflected back and lost, $1/S$, is small. This can be considered as a small channel loss. This loss is cumulative and can limit the total number of users. At each multiplexer, the wavefunction of the incoming photons is spread with a different code. The parts of the wavefunction that end in the reflected part of the spectrum is different for each added photon and, when the signal is restored at the second modulation, we can consider the total effect as a uniform loss.

\section{Demultiplexing}
\begin{figure}[htbp]
\centering
\includegraphics{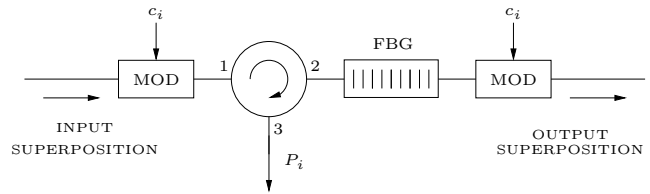}
\caption{Demultiplexer: A receiver extracts from a fibre carrying multiple photons the qubit coming from user $U_i$. The input superposition is first multiplied by code $c_i$. This operation despreads photon $P_i$, which is now confined to the part of the spectrum that the fibre Bragg grating reflects. The other photons remain spread. The circulator takes this signal to the grating, which reflects $P_i$ and forwards the other photons, up to a loss $1/S$, to a second modulator which restores their original state. These photons go back to the common optical fibre while the reflected photon $P_i$ can be recovered from port 3 of the circulator.\label{DEMUX}} 
\end{figure}

If we repeat the multiplexing procedure for the $N$ users, we end with a fibre carrying $N$ photons with the qubits of all the users. We can separate them at each of the $N$ intended receivers with the demultiplexing optical circuit of Figure \ref{DEMUX}. Notice that, as the circulator is a non-reciprocal element, we cannot just use the multiplexing circuit in reverse. We need a slight modification in the order of the elements.

In the demultiplexer, the incoming photon superposition is directed to a modulator which spreads the signal with a code $c_i$. The modulator re-spreads all the photons but $P_i$, which is despread. Its wavefunction is now concentrated in its original band of the spectrum $W$. 

The new signal is then sent through the circulator into a fibre Bragg grating. The grating reflects the desired photon back to the circulator which directs it to its intended recipient (along with a residual noise signal with a fraction $1/S$ of the wavefunction of the other photons).  

The largest part of the wavefunctions of the previous photons goes through the grating and meets a second modulator which multiplies the signal again by $c_i$ and restores the original superposition minus the extracted photon. We can repeat the procedure $N$ times to extract all of the incoming photons.

\section{Simulation}
A practical implementation of the proposed system will introduce new variables that make photon losses larger than the ideal $1/S$ approximation. In this section, we simulate a concrete realization of our add-drop architecture to quantify additional effects. We will see that, although there are higher losses and crosstalk between users, the general behaviour of the system is as expected, with a better performance for longer codes. 

\subsection{Photon shape}
We have simulated a multiplexing scheme for single photons sent in time bins of a length $T$. Each bin can either be empty or carry a Gaussian wave packet. We have chosen this model because it is closely related to two existing quantum key distribution systems, those that use the coherent one-way, COW, protocol \cite{SBG05,Stu09} and Ekert-style protocols \cite{Eke91} with time-energy entangled photons \cite{TBZ00}. 

In COW protocols, logical $\ket{0}$ and $\ket{1}$ values of a qubit are encoded in the time of arrival of a pulse with an average photon number smaller than one. There are two time bins of the same length. A pulse in the first bin corresponds to $\ket{0}$ and a pulse in the second bin to $\ket{1}$. Our example with Gaussian pulses in discrete time bins can describe a COW system. 

In time-energy quantum key distribution protocols, the most important task is distributing to each party one of the two entangled photons coming from a spontaneous parametric down-conversion process. Each of the photons from the entangled state has a wavefunction that can be modelled as a Gaussian \cite{FMV09}. The multiplexing scheme we simulate can describe a delivery system that takes those entangled pairs to multiple users. 

In the simulation, we have used Gaussian wave packets centered in the middle of the time bin and with a width given by a standard deviation of $\sigma=0.1T$. The system has been tested for both wave packets that are in phase and for pulses with a random global phase for each user. The general results are similar in both cases.

\subsection{Codes}
A key element in spread spectrum systems is the choice of codes. For this example, we have generated our codes with linear feedback shift registers (LFSR). This kind of system is of widespread use due to its simplicity. A basic electronic system with shift registers and XOR gates is enough to provide long pseudo-random binary sequences. In particular, it can be shown that, with the right feedback, an LFSR with $n$ registers can produce a periodic output of length $2^n-1$ known as an $m$-sequence. These $m$-sequences have many desirable properties. In one period, they reproduce the statistics of random signals and they have low autocorrelation \cite{GG04}. We can use the shifted versions of an $m$-sequence as our codes. The feedback configuration used to ensure the output of the LFSR is really an $m$-sequence has been taken from the table in \cite{Mut96} and the initial state of each shift register was chosen to be 1. User $i$ is assigned a code $c_i$ that is a circular shift of the initial code by $i$ positions. If we take binary $\pm 1$ data and codes with indices from 1 to $2^n-1$, the inner product of two codes is $c_i\cdot c_j^T=-1$, if $i\neq j$, and $c_i\cdot c_j^T=2^n-1$, if $i=j$.

With the selected codes, we need to assume the multiplexer and demultiplexer blocks can be synchronized to recognize the beginning of each time bin. A large enough time shift can result in a change of code and interference. Other code families might have different sensitivities to time shifts.

We only need synchronization between the modulators. If all the add and drop nodes agree on when a time bin starts, the codes will remain almost orthogonal. We can use a classical side channel to coordinate the nodes. There are working examples of precise synchronization methods in existing quantum key distribution networks \cite{SBG05,TFN08} and we can apply them to our scheme.

\subsection{Filters and modulators}
The filter of the FBG has been modelled as a Gaussian filter with a spectral width $\sigma_{filt}$ that is $\frac{8\pi}{5}$ times the spectral width of the Gaussian wave packet of the photons. The bandwidth has been chosen to be wide enough to reflect most of the desired photon but also to be as narrow as possible in order to minimize losses and prevent the photons of other users to reach the wrong receiver. 

The Gaussian shape approximates the transfer function of apodized gratings. We have also tested additional transfer functions and have obtained results that are qualitatively similar to the ones we present.
 
We have supposed an ideal modulator with abrupt transitions that introduces either a $0$ or a $\pi$ phase shift. Each binary value of the code determines the phase shift the modulator applies in an interval of length $T/S$. The code elements $1$ or $-1$ are generated at regular intervals. After $T$ seconds we have $S$ elements. The modulator applies no phase shift when the code element is $1$ and a $\pi$ shift when the current code element is $-1$.

\subsection{Insertion losses and noise}
In the simulation, we assume the circulators, modulators and all the connections are ideal and have no losses. The aim of this simulation is to model intrinsic losses due to the spreading. The effect of other sources of loss is commented in the discussion. Likewise, we have not included noise or other effects that can degrade the signal. 

\subsection{Results}
The presented multiple access system has two main sources of error: photon loss and crosstalk. The behaviour of the system depends on the spreading factor $S$ and the number of users. Larger spreading factors improve the overall performance. Each additional user introduces degradation to the other channels. 

\begin{figure}[htbp]
\centering
\includegraphics{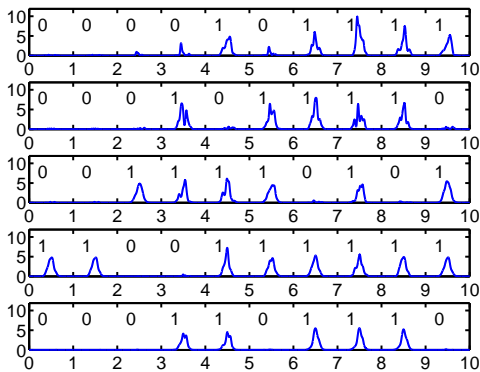}
\caption{Example of transmission in a system with $S=2^8-1$ (with a code from an LFSR with $n=8$ registers) and five users. The graph shows a proxy for $|\psi(t)|^2$ in normalized time $t/T$. The output shows there are losses (low amplitude 1 pulses) and crosstalk (pulses in 0 bins and too high amplitude pulses in 1 bins).\label{pulsesn8}} 
\end{figure}

\begin{figure}[htbp]
\centering
\includegraphics{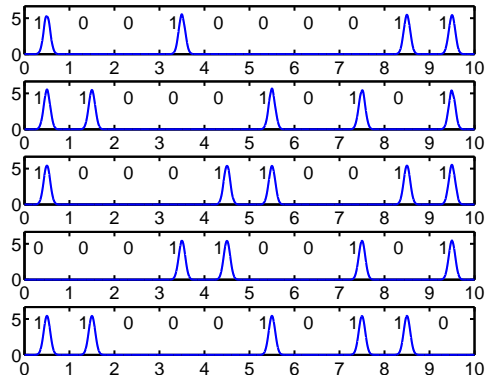}
\caption{Example of transmission in a system with $S=2^{15}-1$ (with a code from an LFSR with $n=15$ registers) and five users. The graph shows a proxy for $|\psi(t)|^2$ in normalized time $t/T$. For this value of $S$, the output reproduces with good accuracy the input state.\label{pulsesn15}} 
\end{figure}

Figures \ref{pulsesn8} and \ref{pulsesn15} show a typical output for spreading factors $S=2^8-1$ and $S=2^{15}-1$ respectively. The simulated system has five users that send a random sequence with eight bits, where a 0 corresponds to an empty time bin and a 1 to a Gaussian pulse with one photon. Time has been normalized to the bin length $T$. The Figures show the output in terms of the density of the average photon number. The area below each pulse in a bin gives the average photon number found during that time $T$ at the corresponding receiver. In an ideal system, the represented output corresponds to the probability density squared $|\psi(t)|^2$. The simulated system includes losses (there can be less than one photon in an occupied time bin) and crosstalk (there is some probability of finding a photon in a supposedly empty time bin or more than one photon in a bin where there should be only one).

Figure \ref{pulsesn8} can help to illustrate both problems. It shows how pulses can be distorted during the multiplexing process. It also has bins labelled with 1 with pulses of different peak heights, which means there are pulses that suffer losses. Apart from that, there are residual pulses in the bins labelled with 0 where part of the photons in other channels reach a user that should find zero photons.   

Figure \ref{pulsesn15} shows how, for a high enough value of $S$, we can recover almost perfect transmission. There is no appreciable distortion, losses or crosstalk. In this case, the density of the average photon number we represent gives a close approximation to $|\psi(t)|^2$. The area under each Gaussian pulse is close to 1 (there are almost no strange photons and losses are low). 

In order to show the effect of the spreading factor and the number of users in the probability of photon loss or the probability of a photon entering the wrong channel, we have performed additional simulations. 

The probability of photon loss has been computed for a communication scenario where only one channel sends a Gaussian single photon pulse and the rest of the users do not transmit. The occupied channel has been randomly chosen from all the available channels. Table \ref{Ploss} shows the average photon loss probability for 200 tests. The Table gives the probability of finding the photon in its original channel as computed from our approximation to the probability density. 

\begin{table}[!t]
\caption{Probability of photon loss}
\label{Ploss}
\centering
\begin{tabular}{|c|c|c|c|}
\hline
 & 5 users & 20 users & 50 users\\
\hline
$S=2^8-1$ \phantom{\big{[}}  &  0.3240  &  0.8301  &  0.9893 \\
\hline
$S=2^{10}-1$ \phantom{\big{[}} &   0.1199  &  0.3723  &  0.6729 \\
 \hline
$S=2^{12}-1$ \phantom{\big{[}} &   0.0585  &  0.1339  &  0.2642 \\
 \hline
$S=2^{14}-1$\phantom{\big{[}}&   0.0426  &  0.0620 &   0.0998 \\
\hline
\end{tabular}
\end{table}

Photon losses increase with the number of users. The photon must cross more multiplexers and demultiplexers, each of which introduces new losses in the filtering stage. Filtering becomes more and more selective for larger spreading factors. As the bandwidth of the spread photons increases, there is less residual loss. We can see that, for a large enough value of $S$, we can bring down the losses to acceptable levels.

A second concern is crosstalk. We have computed the probability of a photon appearing in an empty channel by simulating a situation in which all the channels but one send a single photon pulse with a random global phase. We have randomly assigned the empty channel and checked the probability of having a photon at its output. Table \ref{Px} shows the average probability of crosstalk for 128 runs where each user sends 8 bits. The values in the Table have been computed by integrating the average photon number density in the channel that should be empty.  

\begin{table}[!t]
\caption{Crosstalk probability}
\label{Px}
\centering
\begin{tabular}{|c|c|c|c|}
\hline
 & 5 users & 20 users & 50 users\\
\hline
$S=2^8-1$ \phantom{\big{[}}& 0.0634 & 0.2244 &  0.3889\\
\hline
$S=2^{10}-1$ \phantom{\big{[}} & 0.0185 & 0.0730 & 0.1679\\
\hline
$S=2^{12}-1$ \phantom{\big{[}} &   0.0043 &  0.0186  & 0.0483\\
\hline
$S=2^{14}-1$ \phantom{\big{[}} &  0.0010 & 0.0050   & 0.0127\\
\hline
\end{tabular}
\end{table}

As expected, with more channels there appear more photons and the probability of crosstalk increases. We can also see that a longer code helps to isolate channels better. In any case, crosstalk can be brought to reasonable levels in practical scenarios with many users with an adequate choice of codes. 

\begin{figure}[htbp]
\centering
\includegraphics{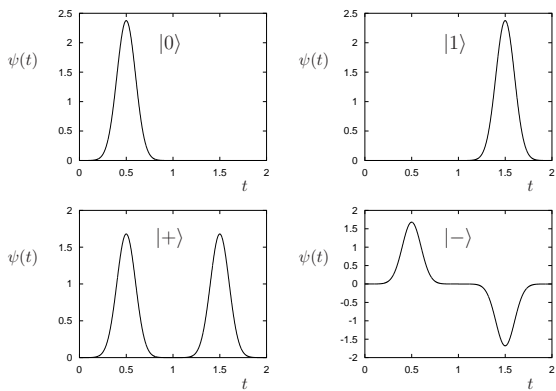}
\caption{Wavefunction of the photon states of a COW quantum key distribution system. The graphs show $\psi(t)$ in normalized time $t/T$.\label{states}} 
\end{figure}

If we use the system for quantum communication, we must also show it can faithfully preserve quantum superpositions. We discuss a simple example that can be found in real-world quantum key distribution, QKD, systems such as coherent one-way, COW, quantum key distribution networks \cite{SBG05}. In the example, we have four possible quantum states. We use two time bins of length $T$. A state with a photon in the first bin and an empty second bin is labelled as $\0$ and we call $\1$ to a state with a photon in the second bin and an empty first bin. We also have two states $\ket{+}=\frac{1}{\sqrt{2}}\left(\0+\1\right)$ and $\ket{-}=\frac{1}{\sqrt{2}}\left(\0-\1\right)$ which are superpositions of the $\0$ and $\1$ states and have wavefunctions that span from $0$ to $2T$ (see Figure \ref{states}). 

We consider an initial wavefunction $\psi (t)$ that travels through an optical network so that at the receiver we get a distorted wavefunction $\tilde{\psi} (t)$, ideally as close to $\psi (t)$ as possible. The metric we use to compare the original and the received state is the fidelity, $F$, defined from the overlap integral,
\begin{equation}
F={\left| \int \tilde{\psi}^{*}(t) \psi(t) \,dt \right|}^{2},
\end{equation}
where we have taken the wavefunction $\psi (t)$ of the original photon at the point of insertion and the wavefuntion $\tilde{\psi} (t)$ of the same photon at the receiver and have discounted the effect of the time of flight through the network. 

Table \ref{fidelity} shows the average fidelity between the input and output photons for the four relevant states in a system with $S=2^{10}-1$. In all the cases the fidelity is close to 1. The Table gives the complementary value $1-F$ which allows a better comparison. 

In the simulation, the modulator applies the code once in each time bin of length $T$. We estimate the effect of filter distortion by sending a photon in a state chosen at random from the four options through an otherwise empty network with random insertion and extraction points. Table \ref{fidelity} presents the fidelity between the input state and a normalized output state. The results tell how similar are the expected and the actual photon states when we do find a photon.

\begin{table}[!t]
\caption{Fidelity (1-F)}
\label{fidelity}
\centering
\begin{tabular}{|c|c|c|c|}
\hline
 & 5 users & 20 users & 50 users\\
\hline
$\0$ \phantom{\big{[}}& $1.079 \cdot 10^{-3}$ & $2.340 \cdot 10^{-3}$ &  $5.626 \cdot 10^{-3}$\\
\hline
$\1$ \phantom{\big{[}} & $1.079 \cdot 10^{-3}$ & $2.340 \cdot 10^{-3}$ &  $5.623 \cdot 10^{-3}$\\
\hline
$\ket{+}$ \phantom{\big{[}} & $1.079 \cdot 10^{-3}$ & $2.342 \cdot 10^{-3}$ &  $5.632 \cdot 10^{-3}$ \\
\hline
$\ket{-}$ \phantom{\big{[}} & $1.078 \cdot 10^{-3}$ & $2.334 \cdot 10^{-3}$  &  $5.606 \cdot 10^{-3}$\\
\hline
\end{tabular}
\end{table}

In general, the effect of the multiplexing and demultiplexing stages is small, but, as expected, we see that the quality of the states degrades as the number of users grow. 

\section{Discussion}
We have proposed a multiple access scheme that brings code division multiple access into the quantum realm. We take $N$ qubits encoded into $N$ separate photons and send them together using the same optical fibre. 

The wavefunction of each photon is spread using a code unique to each user. If the code has $S$ distinct elements ($S$ chips), a wavefunction of bandwidth $W$ is stretched to a bandwidth $SW$. 

At each multiplexing stage, the qubit of a new user is added to the photons already in the channel. During the procedure, the ``old'' photons have a probability $1/S$ of being lost, even for ideal elements. The same happens at each demultiplexing stage. If $N$ users share the channel, a photon can at most undergo $2N-2$ lossy multiplexing/demultiplexing stages. There is a maximum probability of photon loss $\frac{2N-2}{S}$, which can be reduced if the qubits of each user are added and extracted at the right points. We can, for instance, require that the first photon in is the first photon out. In practice, the optical circuits of Figures \ref{MUX} and \ref{DEMUX} will introduce coupling losses that should also be taken into account.

In any case, the system seems adequate for Quantum Key Distribution in its present form. All the necessary elements are already used in optical fibre systems and the codes can be assigned using existing CDMA schemes. Ideally, a large value of $S$ is desirable. It both reduces the unavoidable $1/S$ loss at each element and increases the number of potential users. For $N$ users with $N\leq S$, there are enough orthogonal codes to allow for perfect separation, but it is possible to go above $S$ users using nearly orthogonal codes with a small overlap. A recent experiment on spread spectrum modulation of a single photon shows that a value of $S$ around $2^{15}-1$ is technically feasible \cite{BCY10}. However, there is a practical limit to the value of $S$. Current modulators can achieve modulation rates around 10-100 Gbps, but modulators working at their fastest rate produce smoother transitions in the codes. In our system, simulations show that smooth transitions degrade the overall performance of the multiplexing scheme. This puts an upper bound on the modulation rate and the length of the code. With reasonable modulation rates, a value of $S$ around $2^{13}-1$ or $2^{14}-1$ can only be obtained for photons with wavefunctions in the microsecond length range (and photon rates in the order of MHz). High speed QKD systems aim to achieve photon rates close to the GHz range. The presented spread spectrum system sacrifices photon rate in order to have a flexible and convenient add-drop multiple access scheme.  

Coupling losses at the optical elements are also likely to be a major limitation. In many QKD networks, losses limit the maximum communication distance, but spreading provides certain protection against noise that reduce the impact of losses on the signal-to-noise ratio of the data link. The noise that has been picked up in the channel is spread at the receiver and the filter that rejects adjacent channels also stops a greater proportion of the energy of the noise. This is an independent effect of spreading and can be used to extend the reach of QKD links with a single photon. In that case, the photon needs not to be spread with a modulator. An interesting alternative is using spread spectral teleportation, a teleportation protocol that can stretch or shrink the wavefunction in frequency \cite{Hum10}. This kind of teleportation could extend the applicability of spread spectrum methods to quantum repeater networks \cite{BDC98}.

If losses remain a problem, we can, anyway, limit the number of users and still choose as large an $S$ as permitted by the modulation (the speed at which we can change the phase) to reduce the $1/S$ losses. 

An alternative way to limit losses is using integrated optical elements. For instance, microring structures can replace the fibre Bragg gratings and the circulators and act as frequency selective filters and routing devices \cite{LFS98,XKS08}. The whole multiplexing and demultiplexing subsystems can thus be integrated into one compact optical element. A detailed analysis of this microring-based solution with a full simulation of losses will be presented elsewhere.

The proposed multiplexing method can also be used to combine classical and quantum data. Most quantum key distribution networks send photons through what are called dark fibres, which are reserved for quantum use and carry no classical data. Classical and quantum information channels can share the same fibre if they are assigned different frequency bands, but Raman scattering and other processes triggered by the classical optical signal introduce noise into the photon channel. The proposed add-drop architecture offers a new way to introduce a single photon into an optical fibre that carries classical signals. The method allows insertion in already deployed optical networks and spreading helps to fight the noise the classical signal introduces into the quantum channel. 

As a final note, we would like to remark that, although the presented multiple access scheme has been designed with single photon channels in mind, both the discrete elements and integrated microring versions of the system are also an interesting alternative to classical multiple access techniques. In particular, the low losses of the scheme can eliminate or at least mitigate the need for amplifiers and give energy efficient optical CDMA systems.

\section*{Acknowledgment}
This research has been funded by MICINN project TEC2010-21303-C04-04.

\newcommand{\noopsort}[1]{} \newcommand{\printfirst}[2]{#1}
  \newcommand{\singleletter}[1]{#1} \newcommand{\switchargs}[2]{#2#1}

\end{document}